\documentstyle[aps,epsfig,prd,eqsecnum,preprint]{revtex} 
\tightenlines
\def \beq{\begin{equation}}         \def \eeq{\end{equation}}
\def \beqa{\begin{eqnarray}}        \def \eeqa{\end{eqnarray}}
\def \bea{\begin{array}}	    \def \eea{\end{array}}
\def\npb#1#2#3{    {\it Nucl. Phys. }{\bf B#1} (19#2) #3}

\def\plb#1#2#3{    {\it Phys. Lett. }{\bf B#1} (19#2) #3}
\def\prd#1#2#3{    {\it Phys. Rev. }{\bf D#1} (19#2) #3}

\def\prl#1#2#3{    {\it Phys. Rev. Lett. }{\bf #1} (19#2) #3}
\def\ijm#1#2#3{    {\it Int.J.Mod.Phys.}{\bf A#1}  (19#2) #3}
\def\mpla#1#2#3{   {\it Mod. Phys. Lett. }{\bf A#1} (19#2) #3}
\def\zpc#1#2#3{    {\it Zeit. f{\"u}r Physik }{\bf C#1} (19#2) #3}
\def\report#1#2#3{{\it Phys.Rep}{\bf #1}(19#2) #3}
\begin{document}
\title{Separating Different Models from Measuring $\alpha,\ \beta,\ \gamma$  }
\author{{Y.F. Zhou and Y.L. Wu }\\
  {\small Institute of Theoretical Physics, Chinese Academy of Sciences, \\ 
Beijing 100080,  China}}
\maketitle

\begin{abstract}
New physics effects on the $B^0-\bar{B}^0$ mixing and $B$-decay amplitudes are
discussed. By a combined analysis, the models
of new physics can be partially distinguished. 
It is emphasized that the extraction of unitarity angles $\beta$ and $\gamma$
through rare decay $K\rightarrow \pi\nu\tilde{\nu}$  and charged $B$ decay 
$B^{\pm}\rightarrow DK^{\pm}$ respectively is not likely to be affected by new
physics. Such an observation could be used to distinguish the new physics 
effects from different models. For instance, the top quark
two-Higgs doublet model can be easily separated from those models without
new phase in $B$-decay amplitudes, and can be further distinguished from
the most general two Higgs doublet model by its absence of new phase in 
the $B^0-\bar{B}^0$ mixing.
\end{abstract}

\section{Introduction}

  In the standard model(SM) with $SU(2)_L\times U(1)_Y$ gauge symmetry and three
generation of fermions.  The only source of CP asymmetry is a non-zero complex
phase in the Cabibbo-Kobayash-Maskawa(CKM) matrix.  Although SM has been  proved to be
very successful in phenomenology, its accommodation of CP violation through
complex CKM matrix elements has not been seriously tested experimentally.  
At present, CP violation is one of the least understood issues in particle physics,
and is very promising in the search of indications of new physics.

Assuming unitarity of CKM matrix, the phase information of the matrix can be
displayed elegantly by a set of triangles, called Unitarity triangles(UTs) .  The
central topic is therefore the determination of the angles of those triangles.  In
the past years, much efforts have been made in the neutral-Kaon system as well as in
$B^0_d$ and $B^0_s$ mixings.  However, due to large theoretical uncertainties, our
current knowledge of those angles is still very poor.  It is  expected
that in the up coming B-factories, the measurements of time dependent 
CP asymmetries in $B^0$ decays into CP eigenstates will greatly reduce the 
hadronic uncertainties and obtain the precise value of the angles of UT with 
$b$ and $d$ quarks. If those angles are precisely determined, any deviation 
from the SM predictions will clearly signal new physics beyond the SM.

   Supposing all the angles ($ \alpha,\beta,\gamma$) are  completely determined 
through independent measurements, following the analysis in reference \cite{london97},
there are three
distinct ways in which new physics can show up in the measurements of CP
asymmetry, they are:\\  
\indent 1)$\alpha+\beta+\gamma \not = \pi $,\\
\indent 2)$\alpha+\beta+\gamma=\pi $, 
but the value of $\alpha,\beta $ and $ \gamma$ do not agree with the SM
predictions. \\ 
\indent 3) $\alpha+\beta+\gamma = \pi $, $\alpha,\beta $ and $ \gamma$
are consistent with the SM, but measurement of the angles are inconsistent with
the measurements of the side of the UT.
   
   If any one of these three cases really happens in the future B- factory, the new
physics will be established. However,
this only tells us that new physics exactly exists. We still don't know what kind
of new physics is responsible, since there are variety of  models of new physics which can 
affect the value of angles in the same way.  It is of great importance to
distinguish different models of new physics from the experiment. 

The problem of distinguishing various models of new physics has been discussed in 
\cite{london97,nir90}.  It may involve many models of new physics such as 
suppersymmetric models\cite{susy},muti-higgs \cite{weiberg} without 
flavor changing neutral scalars interactions or general two-higgs-doublet model  
with flavor changing neutral scalars interactions (S2HDM) \cite{g-2hdm,wu}, 
left-right symmetric model\cite{lrs},  Z-mediated 
flavor changing neutral currents(FCNC)\cite{zfcnc} and fourth generation\cite{four}. 
The conclusion is that by comparing their contributions  
to $B^0-\bar{B}^0$ mixings and rare leptonic $B$ decays, these models can be partially 
distinguished. If new physics is founded to be the case 1) or 2) as mentioned above.
This would indicate that new physics is probably to be the S2HDM, 
fourth-generation or Z-mediated
FCNC. If new physics is founded through the case 3), the new physics is likely to be
two-Higgs doublet model without  flavor changing neutral scalar interactions 
(such as Model 1 or Model 2) or minimum
suppersymmetry model. This method is very useful, but still not sufficient
to distinguish each of the models especially when several models have similar
effect on the extraction of those unitarity angles. In this paper, We show that 
the combination of analysis on $B^0-\bar{B}^0$ mixing and $B$-decay amplitude via 
time dependent measurement of CP asymmetry \cite{pw} is also an efficient way in 
separating different models. As an example , the top quark two-Higgs doublet model(T2HDM) 
recently discussed in \cite{soni} can be 
distinguished  not only from a large number of model without CP asymmetry in $B$-decay
amplitude, but also from the S2HDM by their different behavior in 
$B^0-\bar{B}^0$ mixing. The paper is organized as follows: In section {\bf I$\!$I} , we present the basic
formulas on distinguishing different models of new physics by considering their difference
in $B^0-\bar{B}^0$ mixing and $B$-decay amplitude, some features of 
the S2HDM and T2HDM as well as their influences on the determination of angles $\alpha, \beta$ and $\gamma$ 
are discussed in section {\bf I$\!$I$\!$I}. The conclusions are presented in  section {\bf I$\!$V}.

\section{basic formulas}
From the theoretical point of view, there are two basic ways in which new physics can
enter the extraction of angles $\alpha,\beta $ and $ \gamma$. One way is via 
$B^0-\bar{B}^0$ mixings, the other is via $B$ decay amplitudes which  is mainly through 
hadronic penguin diagrams, but in some models it may be also through tree diagrams, such as 
the S2HDM and T2HDM. That will be discussed in detail below. 

If $B^0-\bar{B}^0$ mixing is affected by new physics, for example, from additional
heavy particles in the loop instead of $W$-boson, the angle $\beta$ measured in
the process $B^0_d\rightarrow J/\psi K_S$ can be largely modified. In the SM the time
dependent  asymmetry is given as follows:
\beq
Im\lambda =\left(\frac{q}{p}\right)_{B_d}\left(\frac{\bar{A}}{A}\right)
\left(\frac{p}{q}\right)_{K}=-\sin 2\beta,
\eeq
where the term in the first bracket is from $B^0-\bar{B}^0$ mixing which has the value
of $V_{td}V^*_{tb}/V^*_{td}V_{tb}$ in the SM, $A(\bar{A})$ is the amplitude of $b\rightarrow c(\bar{c}s)
(\bar{b}\rightarrow \bar{c}(c\bar{s}))$ subprocess, the term in the last bracket is from 
$K^0-\bar{K}^0$ mixing since $K_S$ is involved in the final state.  
Without losing generality,
the new  physics can affect  all these three quantities. Let us denote $\phi^{B_d}_{mix},
\phi_A$ and  $\phi^{K}_{mix}$ the new phase from new physics in $B^0_d$ mixing, $B$-decay amplitude,
and $K^0$ mixing respectively. Then the experiment will measure $\beta_{exp}$ 
instead of $\beta_{SM}$ as:
\beq
\beta_{exp}=\beta_{SM}+\phi^{B_d}_{mix}+\phi_A(b\rightarrow c) + \phi^{K}_{mix}
\label{beta}
\eeq
where the process $b\rightarrow c$ indicated in the bracket is a 
tree level transition. The angle $\alpha$ can be measured from $B^0_d$ decay to 
$\pi^+\pi^-$. In this channel there are also contributions from penguin diagrams with 
different strong phase, this can be eliminated by using the isospin analysis\cite{alpha}.
  As it was pointed out in\cite{nir90} and 
also emphasized by many authors\cite{fleisher97,london9702,nir91}, if the angle $\alpha$ is measured through 
the decay channel $B^0_d\rightarrow \pi^+ \pi^-$, then the new physics effect will 
give contributions with an opposite sign as follows:
\beq
\alpha_{exp}=\pi-\beta_{SM}-\gamma_{SM}-\phi^{B_d}_{mix}-\phi_A(b\rightarrow u)  
\eeq

Here the process is ($b\rightarrow u$) as $B^0_d\rightarrow \pi^+\pi^-$ is 
dominated by $b\rightarrow u$ transition.
Consequently, the new phase $\phi^{B_d}_{mix}$ in $B^0-\bar{B}^0$ mixing $cancels$ 
each other in the sum $\alpha_{exp}+\beta_{exp}$. When the new phase from amplitudes
and $K$-meson mixing are  negligible small( these happens in many models
), the sum  $\alpha+\beta$ will remain
unchanged. If the angle $\gamma$ is determined through charged $B$ decay 
$B^{\pm}\rightarrow D K^{\pm}$, since $B^0$-mixing is absent and  the FCNC will not 
be involved in this channel, 
its value can hardly be modified by new physics, so $\gamma_{exp}$ is likely to be 
unchanged and equals $\gamma_{SM}$. 
Thus the sum 
\beq
\alpha_{exp}+\beta_{exp}+\gamma_{exp}=\pi \label{pi}
\eeq
still holds as in the case of SM.

The new phase $\phi^{K}_{mix}$ in $K$-mixing is often thought to be  small, 
this is because the extremely small values of $\Delta m_K$ and $\epsilon_K$  impose a 
very strong constraint on the contributions to $K^0-\bar{K}^0$ mixing from new physics.
Thus as a consequence, the new physics can not produce a relative large value of 
$\phi^{K}_{mix}$. It is a possibility that by observing the violation of equation:
\beq Im\lambda(B_d\rightarrow D^+ D^-)=Im\lambda(B_d\rightarrow J/\psi K_S) \eeq 
or \beq Im\lambda(B_d\rightarrow D^+ D^-)=Im\lambda(B_d\rightarrow \phi K_S) \eeq
One is able to probe the new physics in $K^0-\bar{K}^0$ mixing \cite{nir91}.
In the following discussion we always make the assumption that the new  phase in $K^0$ mixing
is negligible.

Although the effect of decay amplitude $\phi_A$ is
always thought to be small, its  importance on signaling new physics should
not be neglected. As being stressed in reference \cite{grossman}, the effects
of new physics in decay amplitudes are manifestly non-universal, because
they strongly depend on the specific process and decay channel under consideration. 
On the other hand, the effects on $B^0-\bar{B}^0$ mixing are almost insensitive
to the decay modes. Since in general $\phi_A(b\rightarrow c) \not = \phi_A(b\rightarrow u)$
, in the condition that $\phi^{K}_{mix}$ is zero, the equation(\ref{pi})
becomes
\beq
\alpha_{exp}+\beta_{exp}+\gamma_{exp}=
\pi+\phi_A(b\rightarrow c)-\phi_A(b\rightarrow u). \label{new}
\eeq
this will be a clear signal of new physics from decay amplitude.

  By considering whether the equation(\ref{pi}) holds, the models of new physics can
be cataloged into two classes, i.e. models with or without new phase in decay amplitude.
A large number of models such as the 2HDM of types {\bf I} and {\bf I$\!$I},  left-right
symmetric, and the minimum suppersymmetric model, fourth generation and  Z-mediated 
flavor changing neutral currents(FCNC) fall into the first class\cite{london97}, whereas 
the S2HDM and T2HDM as well as some other models fall into the second class. 
Therefore the models can be partially distinguished in this way.

 Another useful information is the new phase in $B^0-\bar{B}^0$ mixing. However, due to
their  cancellation in the sum $\alpha+\beta$ it can not be extracted directly. If one
looks at the $B^0_d$ decay to CP eigenstates such as $J/\psi K_S, D^+D^-$ or $\phi K_S$
the new phases from $B^0$-mixing and from $B$ decay amplitude may mix with each other,
and the final result is the sum of these two kind of contributions. Thus the extraction
of pure new phase from $B^0-\bar{B}^0$ mixing in $B^0_d\rightarrow J/\psi K_S$ becomes
difficult. 

This situation can be simplified if one of the decay amplitudes $\phi_A(b\rightarrow c)$ and
$\phi_A(b\rightarrow u)$ is negligible small. For example, $\phi_A(b\rightarrow u)\sim 0$
while $\phi_A(b\rightarrow c)$ is obviously non-zero ( this happens in the case of the 
S2HDM and T2HDM which is under consideration of this paper ), thus equation(\ref{new})
becomes
\beq 
\alpha_{exp}+\beta_{exp}+\gamma_{exp}=
\pi+\phi_A(b\rightarrow c). 
\label{phia}
\eeq
Thus $\phi_A(b\rightarrow c)$
can be easily obtained by measuring the sum of $\alpha, \beta, \mbox{and } \gamma$. 
Substituting the value of 
$\phi_A(b\rightarrow c)$ into equation({\ref{beta}), the phase $\phi^{B_d}_{mix}$ 
can be fixed. In doing this, we need to know the SM prediction of $\beta_{SM}$
since many decay channels can be seriously polluted by new physics, one should carefully
choose some processes which are not likely to be modified. One way is to use the
value of $|V_{cb}|, |V_{ub}|/|V_{cb}|$ from semileptonic $B$ decays $b\rightarrow 
c(u) \ l \ \bar{\nu_l} $ and $\gamma$ from $B^{\pm}\rightarrow D K^{\pm}$\cite{gronau}. These three    
quantities correspond to two sides and one angle between the two sides in UT.
Thus the whole triangle including angle $\beta$ can be completely determined. 
The shortcoming here is that the prediction of $|V_{ub}|/|V_{cb}|$  is model
dependent and suffer a large theoretical uncertainties.

An alternative way to extract $\beta$ is via rare $K$ decay\cite{buras,buras2,buras3} 
$K\rightarrow\pi\nu\bar{\nu}$. The branching ratio of decay $K^+\rightarrow\pi^+\nu\bar{\nu}$ 
is given as follows:
\beq
B(K^+\rightarrow\pi^+\nu\bar{\nu})=
 \kappa \left[ \left(\frac{Im\lambda_t}{\lambda^5} 
 X(x_t)\right)^2 + \left(\frac{Re\lambda_c}{\lambda} P_0(K^+) + \frac{Re\lambda_t}
{\lambda^5}X(x_t)\right)^2 \right]
\label{k1}
\eeq 
with
\beq
\kappa=\frac{3 \alpha^2 B(k^+\rightarrow \pi^0 e^+ \nu)}
{2 \pi^2 \sin^4 \theta_W} \lambda^8=4.64\times 10^{-11} 
\label{k2}
\eeq
where $X(x_i)$ is an integral function given in \cite{buras}, $x_t=m_t^2/m_W^2,  
\lambda_i=V^*_{is}V_{id}$ and $\lambda=|V_{us}|\sim 0.22$. The function
$P_0(K^+)$ has the form $P_0(K^+)=(2X^e_{NL}/3+X^{\tau}_{NL}/3)/\lambda^4$.
By combining the branching  ratio of $K_L\rightarrow \pi^0 \nu\bar{\nu}$:
\beqa
B(K_L\rightarrow \pi^0 \nu\bar{\nu}) &=&\kappa_L \left(\frac{Im\lambda_t}{\lambda^5}
X(x_t)\right)^2 \\
\kappa_L=\kappa\frac{\tau(K_L)}{\tau(K^+)}&=&1.94\times 10^{-10}
\eeqa  
one can find:
\beq
Im\lambda_{t}=\lambda^5 \frac{\sqrt{B_2}}{X(x_t)} \ \ \ \ 
Re\lambda_{t}=-\lambda^5 \frac{\frac{Re\lambda_c}{\lambda}P_0(K^+)+\sqrt{B_1-B_2}}{X(x_t)}
\eeq
where $B_1$ and $B_2$ are the reduced branching ratios with 
$B_1=B(K^+\rightarrow \pi^+\nu\bar{\nu})/4.64\times 10^{-11}$ and  
$B_2=B(K^+\rightarrow \pi^0\nu\bar{\nu})/1.94\times 10^{-10}$.
 
Using the standard parameterization of the CKM matrix, the angle $\beta$
can be determined by
\beq
\sin 2\beta=\frac{2r_s}{1+r^2_s}
\eeq
with $r_s=(1-\bar{\rho})/\bar{\eta}$. The parameter $\bar{\rho}$ and $\bar{\eta}$ is given 
as follows:
\beqa
\bar{\rho}&=&\frac{\sqrt{1+4s_{12}c_{12} Re\lambda_t/s^2_{23}-(2s_{12}c_{12}
Im\lambda_t/s^2_{23})^2}-1+2s^2_{12}}{2c^2_{23}s^2_{12}} \\
\bar{\eta}&=& \frac{c_{12} Im\lambda_t}{s_{12}c^2_{23}s^2_{23}}
\eeqa
It is well known that the rare decays $K^+\rightarrow \pi^+\nu\bar{\nu}$ and 
$K_L\rightarrow \pi^0\nu\bar{\nu}$ can be calculated with smaller theoretical
uncertainties. These uncertainties can be further reduced in the next-to-leading
 order QCD corrections\cite{NLO1,NLO2,NLO3}  . As a result, the measurement of both
 two decays with an error of $\pm10\%$ will yield $\sin 2\beta$ with an 
 accuracy comparable to the determination from CP asymmetry in $B$-decays
 prior to LHC\cite{buras}.  

Here we emphasize that these channels are not likely to be affected by new 
physics models, especially, the models with FCNC.
This is because the couplings between fermions and additional scalars which often present in
the models of new physics are proportional to the fermion mass, 
the decay involving leptons in the final states will greatly suppress the 
tree level contributions from those scalars. Although there may be significant 
new physics contributions to $Z$-penguin diagrams,  the angle $\beta$ will remain unchanged
if the new physics do not carry additional new phase. This happens in
many models with additional Higgs bosons, such as 2HDM of type 
{\bf I}, and type {\bf I$\!$I}, minimum supersymmetric models, et.al. 

If $\beta_{SM}$ is extracted in this way, it is then possible to study the 
behavior of different models in $B^0-\bar{B^0}$ mixing independent of 
$B$-decay amplitudes. 
In general, different models have different 
behavior in $B^0$ mixings and decay amplitudes.   
It is therefore possible to 
identify the models by combining the analysis of these two aspects. 
Following this strategy, the T2HDM can be distinguished not only from
those models without new phase in $B$ decay amplitudes, but also from
the S2HDM by its absence of contribution in $B^0$ mixing.
 This will be further discussed in the next section.

\section{ On S2HDM and T2HDM }
Let us briefly present some important prospects of T2HDM.
The Lagrangian of T2HDM is as follows:

\beq
{\cal L}_Y=-\bar{L}_L \phi_1 E l_R - \bar{Q}_L \phi_1 F d_R 
    - \bar{Q}_L \tilde{\phi}_1 G I^{(1)} u_R 
    - \bar{Q}_L \tilde{\phi}_2 G I^{(2)} u_R + h.c,
\eeq    

Where $\bar{L}_L$ and $\bar{Q}_L$ are the ordinary left-handed lepton and
quark doublets, $u_R$ and $d_R$ are right-handed singlet quarks,
 $\phi_1$ and $\phi_2$ are two Higgs doublets with $\tilde{\phi}_i=i 
\sigma^{2} \phi^*_i$ and $E,F,G$ are Yukawa coupling matrix. 
$I^{(1)}$ and $I^{(2)}$ are two diagonal matrix with $I^{(1)}_{ij}=\delta_{ij}
(i,j=1,2)$ and $ I^{(2)}_{ij}=\delta_{ij} (i=j=3)$.

Comparing with other quarks, the top quark is in a special status in this model,
i.e. Only $\phi_2$ couples to $t_R$. Let the vacuum expectation value(VEV) of two Higgs
fields to be $v_1/\sqrt{2} $ and $v_2 e^{i \delta}/\sqrt{2}$ respectively. 
If we choose the ratio between two
VEVs $\tan\beta=|v_2|/|v_1|$ to be large ($\tan\beta $ is close to $m_t/m_b$), 
the large mass of top quark  can be naturally explained. That is the motivation 
of proposing this model. 

The charged quark-Higgs Yukawa interaction in this model reads
\beqa
%L^N & = & -\sum_{i=u,c,t} m_i \bar{u}^i_L u^i_R \frac{{\phi^0}^*_1}{v_1^*}
%    - \sum_{i,j=u,c,t} \bar{u}^i_L \Sigma_{ij} u^j_R 
%     \left( \frac{{\phi^0}^*_2}{v^*_2}-\frac{{\phi^0}^*_1}{v^*_1} \right) +h.c\\
{\cal L}^C =&-& 2\sqrt{2} G_F [ -\bar{u}^i_L V_{ij} m_{d_j} d^j_R \tan\beta \nonumber\\
           &+& \bar{u}^i_R mu_i V_{ij} d^j_R \tan\beta
                     +\bar{u}^i_R \Sigma^{\dagger}_{ij'} V_{j'j} d^j_L(\tan\beta+\cot\beta)
                    ] H^+ +H.c,
\eeqa     
where $m_i$ are the quark mass eigenstates, V is the usual CKM matrix. 
The matrix $\Sigma$ can be parameterized as follows: 

\beq
\Sigma=	\left(\bea{ccc}
          0 &    0     & 0 \\
          0 & m_c \epsilon_{ct}^2 |\xi |^2  & m_c \epsilon_{ct}\xi^* \sqrt{1-|\epsilon_{ct}|\xi |^2}  \\
          0 &   m_c \xi \sqrt{1-|\epsilon_{ct}|\xi |^2} & m_t (1-|\epsilon_{ct}\xi |^2)
          \eea
     \right)	
\eeq

Since there exist none zero off-diagonal elements in $\Sigma$ matrix, this model may lead to flavor
changing neutral currents, but only in up-type quarks. Therefore it will enhance the $D^0-\bar{D}
^0 $ mixing provided that $\tan\beta$ is large\cite{kao,soni}. On the other hand, it has little effect on 
$K^0-\bar{K}^0$ mixing and $B^0-\bar{B}^0$ mixing since these mesons contain  down-type quarks.  
Another distinct feature is that the off-diagonal element $\Sigma_{32}$ can be quit large,
thus the model generally has a very large couplings for the vertex $\bar{b}cH^-$ 
or $\bar{t}cH^0$ .

Let us turn to a brief discussion on the S2HDM. This model can be
obtained if we abandon the discrete symmetry which is often imposed on the Lagrangian of 
two Higgs doublet model\cite{model12} and  replac it with an approximate global U(1) 
family symmetry\cite{g-2hdm,wu}. The point is that the  smallness  of the off-diagonal terms 
in the CKM matrix suggests that violation of flavor 
symmetry are specified by small parameters. It then turns out that reasonable choices 
for these small parameters combined with the natural smallness of Higgs boson couplings allows one 
to meet the constraint on flavor changing neutral scalar exchange. 
Since there are no discrete
symmetries, many new sources of CP asymmetry can arise from its 
lagrangian\cite{wu}. Therefore, 
this model can affect the measurement of the angles of UT in 
many different ways. For example the possible large effects on 
$B^0-\bar{B}^0$ mixing\cite{kao}, weak transition $t\rightarrow c$ and
large CP asymmetry in $b\rightarrow s \gamma$ \cite{LWW} have 
been investigated. 

The Lagrangian of S2HDM has the form:
\beq
{\cal L}_Y=\bar{Q}^i \Gamma^U_{1,ij} U_{R_j} \phi_1 + \bar{Q}^i \Gamma^D_{1,ij} D_{R_j} \tilde{\phi_1}
    + \bar{Q}^i \Gamma^U_{2,ij} U_{R_j} \phi_2 + \bar{Q}^i \Gamma^D_{2,ij} D_{R_j} \tilde{\phi_2} + h.c.
\eeq
After a rotation into quark mass eigenstates, it can be rewritten as\cite{wu}:
 
\begin{equation}
{\cal L}_{Y} = (L_{1} + L_{2}) \cdot (\sqrt{2}G_{F})^{1/2}
\end{equation}
with
\begin{eqnarray}
{\cal L}_{1} & = & \sqrt{2} ( H^{+} \sum_{i,j}^{3}
 \xi_{d_{j}} m_{d_{j}} V_{ij} \bar{u}_{L}^{i} d^{j}_{R} - H^{-} \sum_{i,j}^{3}
\xi_{u_{j}} m_{u_{j}} V^{\dagger}_{ij} \bar{d}_{L}^{i} u^{j}_{R} ) \nonumber \\
& & + H^{0} \sum_{i}^{3} (m_{u_{i}} \bar{u}_{L}^{i} u^{i}_{R} +
m_{d_{i}} \bar{d}_{L}^{i} d^{i}_{R} ) \\
& & +  (R + i I) \sum_{i}^{3} \xi_{d_{i}} m_{d_{i}} \bar{d}_{L}^{i} d^{i}_{R} +
(R - i I) \sum_{i}^{3} \xi_{u_{i}} m_{u_{i}} \bar{u}_{L}^{i} u^{i}_{R} + H.c.
 \nonumber
\end{eqnarray}

\begin{eqnarray}
{\cal L}_{2} & = & \sqrt{2} ( H^{+} \sum_{i,j'\neq j}^{3}
 V_{ij'} \mu^{d}_{j'j} \bar{u}_{L}^{i} d^{j}_{R}
- H^{-}\sum_{i,j'\neq j}^{3} V^{\dagger}_{ij'}\mu^{u}_{j'j}
\bar{d}_{L}^{i} u^{j}_{R} )  \\
& & + (R + i I) \sum_{i\neq j}^{3} \mu^{d}_{ij}
\bar{d}_{L}^{i} d^{j}_{R} +
(R - i I) \sum_{i\neq j}^{3} \mu^{u}_{ij} \bar{u}_{L}^{i} u^{j}_{R} + H.c.
 \nonumber
\end{eqnarray}
Where the factors $\xi_{d_{j}}m_{d_{j}}$ arise primarily from diagonal elements
of $\Gamma_{1}$ and $\Gamma_{2}$ whereas the factors $\mu_{jj'}^{d}$ arise from
the small off-diagonal elements.

By abandoning the discrete symmetry, this model obtains rich sources of CP
violation. They can be classified into four major types\cite{wu}: 
  (1) The induced CKM matrix. 
  (2) The phases in the factors $\xi_{f_{i}}$ provide CP violation in
the charged-Higgs exchange processes, which are independent of the CKM phase.
 (3)   The phases in the factors $\mu_{ij}^{f}$. These yield CP violation
in flavor changing neutral scalar interaction.
 (4)   the phase from the mixing matrix of the three neutral Higgs scalars

Although these two models have some similar behavior in CP asymmetry, there still
exist several subtle differences between them.  

First, in the T2HDM the coupling 
between quarks and Higgs boson is determined by only  parameters $\tan\beta$ and $\xi$. 
On the other hand such couplings in the S2HDM are flavor dependent. So that  the latter
has  more freedom in fitting the experimental data.

Second, there is no  complex phase in the diagonal Yukawa couplings in T2HDM. This
means that there is no  CP asymmetry from charged Higgs exchange in $t\rightarrow b$ transition.
So it  will result in a small CP asymmetry in the decay $b\rightarrow s \gamma$ , which is
of the order less than $10^{-2}$. On the contrary in the case of S2HDM, this effect
could be larger\cite{LWW}.  

Third,  although in T2HDM the non-zero complex elements in $\Sigma$ matrix can
lead to FCNC, its effect is constrained only in up-type quarks, there is no FCNC between
down-type quarks. As a result, $K^0-\bar{K}^0$ and $B^0-\bar{B}^0$ mixings 
can not be seriously modified by this model. But they could receive contributions 
in the S2HDM.

 Let us investigate their new physics effects on the 
determination  of angles $\alpha, \beta, \gamma$, respectively .    

The 'gold-plated' channel for determining angle $\beta$ is decay $B^0_d\rightarrow J/\psi K_S$, which
is dominated by tree level $b\rightarrow c$ process. In both S2HDM and T2HDM, 
 a considerable contribution from decay amplitude can arise from $b\rightarrow
c$ transition\cite{soni,KS} . The reason is that the off-diagonal element  $\Sigma_{32}$ can relatively
large in T2HDM. Moreover one can see from equation(3.3) that the CKM matrix elements 
associated with $\Sigma_{32}$
is $V_{tb}$ rather than $V_{cb}$. This can contribute to  an additional enhancement  factor of
$|V_{tb}/V_{cb}|\approx 25$. The effective Lagrangian at tree level $b\rightarrow c\bar{c}s$ has
the form:
\beq
L_{eff}=-2\sqrt{2} G_F V_{cb} V^*_{cs} \left[ \bar{c}_L \gamma_{\mu}b_L \bar{s}_L \gamma^{\mu} c_L
      +2 \zeta e^{i\delta} \bar{c}_R b_L \bar{s}_L c_R \right],
\eeq
where  
\beq
\zeta e^{i\delta}=\left\{ \bea{lc}
              (1/2)(V_{tb}/V_{cb})(m_c \tan\beta/m_H)^2 \xi^*  & \mbox{for \ T2HDM} \\
              \\
              (1/2)(V_{tb}/V_{cb})(\mu^{u\dagger}_{32}/m_H)^2       & \mbox{for \ S2HDM} 
                          \eea
         \right.      
\label{zeta}
\eeq               
Using the formalism in references\cite{soni,KS}. The decay amplitude of $B\rightarrow J/\psi K_S$ can
be written as $A=A_{SM}[1-\zeta e^{-i\delta}]$, where $A_{SM}$ denotes the amplitude in SM.
If factorization holds there is no relative strong phase between $W$-and Charged-Higgs exchange
process. So $B$ and $\bar{B}$ decay only differ by a CP-violating phase. Their ratio is given by:
$\bar{A}/A=(\bar{A}_{SM})/A_{SM}e^{-2i\phi_A}$, where $\phi_A=\tan^{-1}(\zeta\sin\delta/(1-\zeta
\cos\delta))$ is the correction to SM from charged Higgs exchange. Thus the time dependent 
asymmetry will measure $\beta_{SM}+\phi_A$ rather than $\beta_{SM}$
 
 In the case of S2HDM, the new phase from amplitude can be obtained by simply replace
 the expression of $\zeta e^{i\delta}$ in T2HDM in equation(\ref{zeta}).
 However, the situation here is more complicate since there are additional contributions
 to $B^0-\bar{B}^0$ mixing, which will largely change the value of angle $\beta$.
 The new contributions come from the couplings $\xi_i$ and $\mu_{ij}$. They are
 in general complex. Although the value of  $\mu_{ij}$ can be constrained 
 from the measurement in $x_d\equiv\Delta m_B/\Gamma_B$\cite{soni2hdm,kueng,zhou}, 
 due to the large uncertainties of $|V_{td}|$ ,  $\phi^B_{mix}$ can still be rather 
 large\cite{zhou}.

  The decay  $B^0_d\rightarrow\pi^+\pi^-$ is thought
 to be a good channel for the extraction of angle $\alpha$.  Since  $B^0_d\rightarrow
 \pi^+\pi^- $ decay is dominated by  $b\rightarrow u(\bar{u}d)$ tree level process. The additional
 contribution from charged Higgs boson is proportional to the $d$-quark mass $m_d$ which is 
 negligibly small. There are also contributions from charged Higgs loop in penguin diagrams.
 By using the isospin analysis\cite{alpha}, those effects can be eliminated.  
 Thus there are no additional new phases in  $B^0_d\rightarrow\pi^+\pi^-$ decay amplitude.
 In T2HDM, due to the absence of new phase in $B^0-\bar{B}^0$ mixing, this measurement of 
 angle $\alpha$ will give $\alpha_{exp}=\alpha_{SM}$, however in the S2HDM, 
 as has been discussed in the previous section, it will  give 
 $\alpha_{exp}=\alpha_{SM}-\phi^B_{mix}$.

 Finally let us consider the measurements of angle $\gamma$. As being proposed in\cite{gronau}
 , $\gamma$ can be extracted from charged $B$ decay $B^{\pm}\rightarrow D K^{\pm}$. By an 
 independent measurement of six amplitude $B^+\rightarrow D^0 K^+, B^+\rightarrow \bar{D}^0 K^+,
 B^+\rightarrow D^0_{CP} K^+,B^-\rightarrow D^0 K^-,B^-\rightarrow \bar{D}^0 K^-,B^-\rightarrow D^0_{CP} K^- 
 $ the angle $\gamma$ can be in principle determined. Where $D^0_{CP}=\frac{\sqrt{2}}{2}(D^0+
 \bar{D}^0)$ is the CP even eigenstate. 
  In the SM the relations of these six amplitudes are :
 \beqa
 \sqrt{2} A(B^+\rightarrow D^0_1 K^+)&=&A(B^+\rightarrow D^0 K^+)
                                      +A(B^+\rightarrow \bar{D}^0 K^+), \\
  \sqrt{2} A(B^-\rightarrow D^0_1 K^-)&=&A(B^-\rightarrow D^0 K^-)
                                      +A(B^-\rightarrow \bar{D}^0 K^-), 
 \eeqa
 and 
 \beqa
 A(B^+\rightarrow \bar{D}^0 K^+)=A(B^-\rightarrow D^0 K^-), \label{b-to-c}\\
 A(B^+\rightarrow D^0 K^+)=A(B^-\rightarrow \bar{D}^0 K^-).
  \eeqa
 with $|A(B^+\rightarrow D^0_1 K^+)| \not =|A(B^-\rightarrow D^0_1 K^-)|$
 these relations are illustrated in Fig. 1. 
 If the magnitudes of the amplitudes can be measured experimentally, 
 one can then extract the angle $\gamma$.
 
 In both S2HDM and T2HDM,  only $b\rightarrow c$ process could be modified
 considerably by charged Higgs exchange. This implies that 
 the two amplitudes $A(B^+\rightarrow D^0 K^+)$ and $A(B^+\rightarrow D^0 K^+)$
 which are dominated by the $b\rightarrow u$ transitions will remain unchanged and
 the angle between them is still given by $2\gamma$. However, the new phase can contribute to
 $A(B^+\rightarrow \bar{D}^0 K^+)$ and $A(B^+\rightarrow \bar{D}^0 K^+)$ since they 
 are dominated by the $b\rightarrow c$ at tree level subprocess. The relation of (\ref{b-to-c})
 will be modified to be 
 \beq
 A(B^+\rightarrow \bar{D}^0 K^+)= e^{2i\phi_A}A(B^-\rightarrow D^0 K^-)
 \eeq 
 As it is shown in Fig.1, if $\phi_A$ can be extracted from equation(\ref{phia}), 
 then the  angle $\gamma$ can be obtained by measuring those six amplitudes. 
 The angle $\gamma$ determined in this
 way is equal to the one in the SM, i.e. $\gamma_{exp}=\gamma_{SM}$.
 
 In summary, the  $B^0_d\rightarrow J/\psi K_S$ decay
 can be affected in both models. In the T2HDM, it is affected through the $B$ decay amplitudes.
 In the S2HDM , the new phase could arise from both $B^0$ mixing and decay amplitudes. In the 
 extraction of $\alpha$, the effect of T2HDM is negligible, but the one of S2HD can 
 contribute a new phase from decay amplitude through $B^0$ mixing. In the extraction of 
 $\gamma$ , if the new phase from decay amplitude can be determined from equation(\ref{phia})
 , the method of measuring $\gamma$ by combining the six amplitudes of $B^{\pm}\rightarrow
 D K^{\pm}$ still works and $\gamma_{exp}$ will be equal to $\gamma_{SM}$. 
 
 Since $\beta_{SM}$ can be extracted through $K\rightarrow \pi\nu\bar{\nu}$, it is then
 possible to extract $\phi^B_{mix}$ from (2.2) or (2.3). If  $\phi^B_{mix}\not =0$ is observed
 in the future experiment, it implies that the T2HDM is disfavorable.
 
\section{conclusions}
In conclusion, the distinguishment of different new physics models is discussed . It has been seen
that by comparing their different behaviors in $B^0-\bar{B}^0$
mixing and $B$ decay amplitudes, these models can be partially separated. The 
distinguishments between T2HDM and S2HDM have been discussed in detail. 
The new physics effects in measuring UT angles $\alpha, \beta$ and $\gamma$ from 
these two models have been examined.
It has been seen that by measuring $\alpha, \beta, \gamma$ from $B^0_d\rightarrow \pi^+\pi^-,
 K\rightarrow \pi \nu\bar{\nu}$ and $B^{\pm}\rightarrow D K^{\pm}$ respectively, the new phase 
 $\phi^{B_d}_{mix}$ can be extracted. Since there is no contribution from 
 $B^0$ mixing  in T2HDM, if $\phi^{B_d}_{mix}\not =0$ from the future experiment is
 founded, the  T2HDM will be excluded. The situation will be different in the S2HDM where
 a non-zero value of $\phi^{B_d}_{mix}$ is allowed. This is because the T2HDM can be regarded
 as one of the special cases of S2HDM.
 \\
 
 This work was supported in part by the NSF of China under grant No. 19625514.
  
%\begin{thebibliography}

%\end{thebibliography}

\begin{figure}
 \centerline{ \psfig{figure=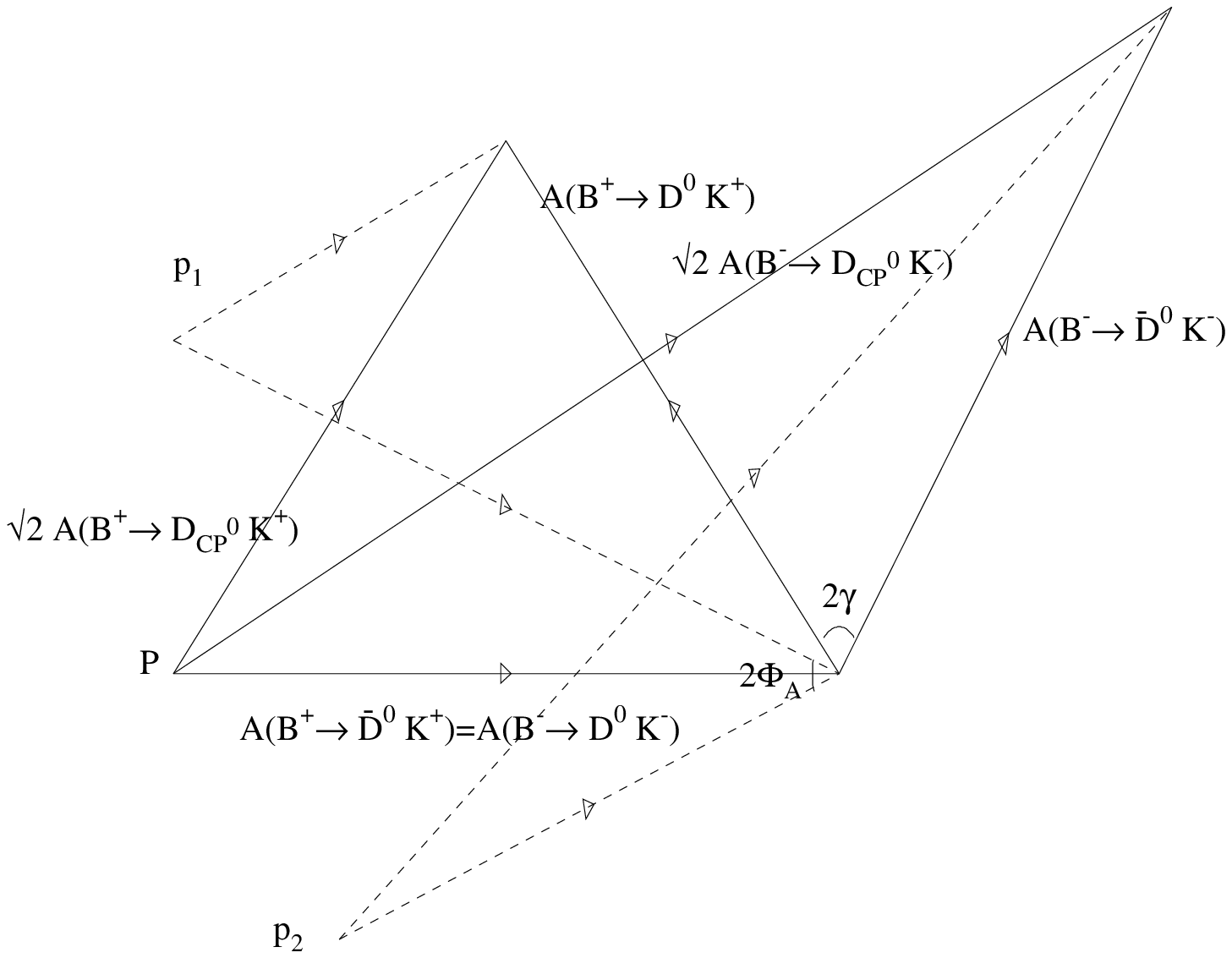, width=15 cm} }
 \caption{
 The triangle relations of six decay amplitudes $B^{\pm}\rightarrow D^0 K^{\pm}
 ,  B^{\pm}\rightarrow \bar{D}^0 K^{\pm},$ and $B^{\pm}\rightarrow D^0_{CP} K^{\pm},$ 
 in the SM(solid line) and in the models with only new phase in $b\rightarrow c$ tree 
 level transition such as S2HDM and T2HDM 
 (dashed line). The relation $A(B^+\rightarrow \bar{D}^0 K^+)=A(B^-\rightarrow D^0 K^-$)
 which holds in the SM is violated when new phase is involved. If the angle
  between them ( denoted by $2\phi_A$) can be determined experimentally, those triangles can
 still be used to extract angle $\gamma$ and will be found to be equal to
 the one in the SM.
 }
 \label{fig1}
\end{figure}

\end{document}